\documentclass[aps,prl,preprint,groupedaddress]{revtex4}

\usepackage{dcolumn}
\usepackage[]{amsmath}
\usepackage[psamsfonts]{amsfonts}
\usepackage{keyval}
\newif
\ifpdf
\ifx\pdfoutput\undefined
\pdffalse 
\else
\pdfoutput=1 
\pdftrue \fi

\ifpdf
\usepackage[pdftex]{graphicx}
\else
\usepackage{graphicx}
\fi

\begin{document}


\newcommand{\beq}{\begin{equation}}
\newcommand{\eeq}{\end{equation}}
\newcommand{\beqar}{\begin{eqnarray}}
\newcommand{\eeqar}{\end{eqnarray}}
\newcommand{\bcen}{\begin{center}}
\newcommand{\ecen}{\end{center}}
\providecommand\ket[1]{| #1 \rangle}\providecommand\bra[1]{\langle
#1 |} \providecommand\sca[2]{\langle #1 | #2 \rangle}
\newcommand\mt[1]{\hat #1}\newcommand\vc[1]{\mbox{\boldmath $#1$}}
\providecommand\du{{\mathrm{d}}}




\title{
Density Functional Theory for the Photoionization Dynamics of Uracil
}

\author{D.~Toffoli }
\affiliation{The Lundbeck Foundation Center for Theoretical Chemistry\\
Department of Chemistry, University of Aarhus, \\
Langelandsgade 140, DK--8000 Aarhus C, Denmark}
\affiliation{ CNR-INFM Democritos, National Simulation Center, Trieste, Italy}
\author{P.~Decleva }
\affiliation{ Dipartimento di Scienze Chimiche,
Universit\'{a} degli Studi di Trieste, Via L. Giorgieri 1, I-34127
Trieste, Italy.}
\affiliation{ CNR-INFM Democritos, National Simulation Center, Trieste, Italy}

\author{
F.~A.~Gianturco }
\affiliation{ Dipartimento di Chimica, Universit\'{a}
degli Studi "La Sapienza", Piazzale A. Moro 5, I-00185 Roma, Italy.
}

\author{
R.~R.~Lucchese } \affiliation{ Department of Chemistry, Texas A\&M
University, College Station, Texas, 77843-3255, USA. }

\date{\today}

\begin{abstract}

Photoionization dynamics  of the RNA base Uracil is studied in the
framework of Density Functional Theory (DFT). The photoionization
calculations take advantage of a newly developed parallel version of
a multicentric approach to the calculation of the electronic
continuum spectrum which uses a set of B-spline radial basis
functions and a Kohn-Sham density functional hamiltonian. Both
valence and core ionizations are considered. Scattering resonances
in selected single-particle ionization channels are classified by
the symmetry of the resonant state and the peak energy position in
the photoelectron kinetic energy scale; the present results
highlight once more the site specificity of core ionization
processes. We further suggest that the resonant structures
previously characterized in low-energy electron collision
experiments are partly shifted below threshold by the
photoionization processes. A critical evaluation of the theoretical
results providing a guide for future experimental work on similar
biosystems.
\end{abstract}


\maketitle

\section{\label{sec:section1}Introduction}

Research on electron- and photon-molecule collision dynamics is of
both fundamental ~\cite{Ehrhardt94} and practical ~\cite{Grill94,
Massey82} importance, for it is needed in a broad variety of
applications, ranging from the manufacturing of semiconductor
devices to atmospheric chemistry and physics. Theoretical studies
are therefore needed not only for a fundamental understanding of the
underlying dynamics, but also for the quantitative  prediction of
the appropriate data.

Following the pioneering work of Sanche and co-workers
~\cite{Sanche00}, electron interactions with biologically important
molecules such as amino acids and nucleotides, have gained
prominence, as it was convincingly suggested that resonant
mechanisms induced by non-thermal low-energy electrons could be
related to DNA and RNA lesions such as single strand breaks (SSBs)
and double strand breaks (DSBs)~\cite{Sanche00}.

Thus, the suggestion that metastable electron capture of the large
amount of non-thermal secondary electrons produced upon the
interaction of ionizing radiation with the cell
environment~\cite{Intnl} by the various DNA components could
initiate the formation of SSBs and DSBs, stimulated a wealth of
experimental work on electron collisions with gaseous~\cite{Hanel03,
Abdoul-Carime00, Denifl04, Feil04} and thin films DNA and RNA bases
~\cite{Penhoat01}. Recently theoretical studies have also given a better understanding of
these insights~\cite{Grandi04, Gianturco04}.

Additionally, the notion of charge transfer and of electron flow along
the $\pi$ stack of nucleobases would benefit from the detailed
knowledge of the response of the nucleotide moieties to excess
negative charge~\cite{Schiedt98}. Thus, the capability to stabilize
an excess of charge in the nucleobases has been extensively
investigated in recent years both experimentally and theoretically,
(see for example refs.~\cite{Desfrancois96, Desfrancois98,
Hendricks96, Hendricks98, Smets96, Smets98, Schiedt98, Aflatooni98,
Wesolowski01, Dolgounitcheva99}) where electron attachment energies
to several nucleobases including the title molecule where evaluated
for the formation of both dipole- and valence-bound anions.

At variance with this promising and stimulating scenario, much less
attention have been paid to the study of photoionization dynamics.
In fact, even if the first measurement of the ionization potential
(IP) of Uracil dates back to the late sixties~\cite{Lifschitz67} and
since then UV and X-ray photoelectron spectra were measured and
interpreted by a large number of investigators (see for
example,~\cite{ODonnell80, Urano89, Kubota96}, and references
therein) none of these studies focused  on the dynamical part of the
photoionization process, namely, on the dependence of the intensity
of the process on the incident photon energy (or, equivalently on
the photoelectron kinetic energy) and on the photoelectron ejection
angle for a given final target state. The first five bands of the
photoelectron spectra have been unambiguously assigned and
associated with the three highest occupied $\pi$ and the two (oxygen
lone-pairs) nonbonding orbitals ($n$), the ordering being $\pi_{1},
n_{1}, \pi_{2}, n_{2}, \pi_{3}$ from the top of the photoelectron
spectrum respectively~\cite{ODonnell80}, and the experimental
ordering has been verified by theoretical semi-empirical and {\em
ab-initio} CI calculations ~\cite{ODonnell80, Urano89, Kubota96} and
is being confirmed by the present work as well as we shall discuss
below. High accurate gas phase ionization energies of Uracil,
obtained by using electron-propagator methods have  been
published~\cite{Dolgounitcheva00} and ionization energy thresholds
in aqueous solutions have been discussed in even earlier
work~\cite{Hernandez04}.

The close connection between shape-resonances in electron-molecule
scattering and those in molecular photoionization has long been
recognized~\cite{Dehmer79a}. In fact, although the long-range part
of the scattering potential is drastically different in the two
cases, the short-range part, is actually quite similar, being
dominated by the interactions between the nuclei and the electrons
common to both collision complexes. Therefore, since shape-resonant
states are localized within the molecular volume, these should
maintain their identity from one system to the other, although
shifted in energy owing to the less effective screening in the
scattering potential of the neutral molecule compared to that for
the positive ion target. As a rule of thumb, therefore,
photoionization shape resonances should have their counterparts in
electron-molecule scattering experiments, albeit shifted by $\sim10$
eV to higher electron energy~\cite{Dehmer87}.

The aim of the present study is to characterize, for the first time
at the DFT level, both the outer valence and inner-shell
photoionization dynamics of the Uracil molecule in a broad energy
range, from threshold up to 100 eV of photoelectron kinetic energy.
The method we use is essentially a one-electron method, thus we are
unable to explicitly account for the variety of many-body effects
inherently displayed in the complexity of electron- and
photon-molecule collisions both at low and high
energies~\cite{Zurales97}, but it has consistently afforded a
realistic description of one-electron resonant processes (for a
review see ref.~\cite{Bachau01}). Further, its LCAO
version~\cite{Toffoli02} is applicable to fairly large molecular
systems with a markedly lower computational cost when compared to
earlier, realistic independent-electron {\em ab-initio}
methods~\cite{Dehmer87}.

The paper is organized as follows: the next two sections describe our
theoretical method and the computational details. In Section IV we
will discuss our results for the core- and outer-valence
ionizations. Our
conclusions and perspectives are then summarized in the final section.

\section{\label{sec:section2}Computational Method}

Dynamical quantities describing the photoionization process from
Uracil are calculated in the framework of the DFT approach, by
employing a Fortan90 suite of codes fully described in earlier
publications~\cite{Toffoli02,Stener04} so we only sketch here their
salient features. The present formulation affords a one-electron
picture of the scattering process, thereby casting the collisional
problem as given via the use of an effective potential according to
the Kohn-Sham scheme ~\cite{Parr89}. The escaping electron is
deflected by such a potential containing the molecular ground state
density, $\rho(\vec{r})$, and conventionally separated into direct
(Hartree), $V_{H}$ and exchange-correlation, $V_{xc}$ terms:

\begin{equation}
\label{eq1} h_{KS}\varphi_{i}=\epsilon_{i}\varphi_{i}
\end{equation}

with

\begin{equation}
\label{eq2}
h_{KS}=-\frac{1}{2}\nabla^{2}-\sum_{i=1}^{N}\frac{Z_{i}}{\left|\vec{r}-\vec{R_{i}}\right|}
+
\int{\frac{\rho(\vec{r'})}{\left|\vec{r}-\vec{r'}\right|}d\vec{r'}+V_{xc}[\rho(\vec{r})}]
\end{equation}

The interaction potential is then expanded in a composite basis set,
whose nature constitute the key-feature of the present LCAO
approach~\cite{Toffoli02}. The LCAO basis set consists in a large
single center expansion (SCE) located at a chosen origin (usually
the center of mass of the molecule)

\begin{equation}
\label{eq3} \chi_{nlh}^{p\mu,
SCE}=\frac{1}{r}B_{n}(r)\sum_{m}b_{lmh}^{p\mu}Y^{R}_{lm}(\theta,\phi)\equiv\frac{1}{r}B_{n}(r)X_{lh}^{p\mu}(\theta,\phi),
\end{equation}
and supplemented by functions of the same type, located on the
off-center arbitrary positions~$j$

\begin{equation}
\label{eq4} \chi_{nlh}^{p\mu, i}=\sum_{j\in
Q_{i}}\frac{1}{r_{j}}B_{n}(r_{j})\sum_{m}b_{lmh,j}^{p\mu}Y^{R}_{lm}(\theta_{j},\phi_{j}),
\end{equation}

In Equation~(\ref{eq4}), index $i$ runs over the non-equivalent
nuclei, $j$ runs over the set of equivalent nuclei, $Q_{i}$, and
gives the origin of the off-center spherical coordinates
$(r_{j},\theta_{j},\phi_{j})$. The sets of coefficients
$b_{lmh}^{p\mu}$ and $b_{lmh,j}^{p\mu}$ define the unitary
transformations between real spherical harmonics
$Y^{R}_{lm}(\theta,\phi)$ and the symmetry adapted angular basis
sets~\cite{Altmann94} which transform as the \emph{$\mu$}th element
of the \emph{p}th irreducible representation (IR) of the molecular
point group. $B_{n}$ is the \emph{n}th spline one-dimensional
function~\cite{deBoor78}. The $B$-splines are built over the radial
interval $[0, R_{max}^{SCE}]$ for the set ${\chi_{nlh}^{p\mu,
SCE}}$, and over the intervals $[0, R_{max}^{i}]$ for the off-center
functions ${\chi_{nlh}^{p\mu, i}}$. In the LCAO implementation the
spheres of radius $R_{max}^{i}$ are not allowed to intersect each
other; furthermore, in order to assure continuity of the second
derivatives over the surfaces of the spheres, for every
${\chi_{nlh}^{p\mu, i}}$ set, the last three splines are deleted.

With this basis set, Eq.~(\ref{eq1}) is recast into an algebraic
eigenvalue problem, and bound state solutions (orbitals
$\varphi_{i}$) are obtained with standard generalized
diagonalization of the hamiltonian matrix, whereas scattering states
are extracted as the set of linearly independent eigenvectors of the
energy-dependent matrix $A^{\dag} A$:
\begin{equation}
\label{eq5} A^{\dag} A(E)c=ac
\end{equation}

corresponding to minimum modulus eigenvalues~\cite{Toffoli02}. In
Eq.~(\ref{eq5}) $A(E)=H-ES$, $H$ and $S$ being the hamiltonian and
overlap matrices over the LCAO basis set respectively.
Diagonalization is efficiently performed with the inverse iteration
procedure \cite{Brosolo92}. Partial-wave independent solutions of
Eq.~(\ref{eq5}) are then normalized to incoming wave boundary
conditions~\cite{Taylor72} and partial cross sections and asymmetry
parameters are evaluated with standard expressions~\cite{Chandra87}.

\section{\label{sec:section3}Computational Details}

The DFT calculations have been executed as follows. The ground state
electron density of Uracil at the experimental equilibrium geometry
~\cite{Stewart67} for its diketo form~\cite{Rejnek05} and assuming a
$C_{s}$ symmetry, was calculated with the ADF package
~\cite{Baerends73,Guerra98} employing an all-electron double-$\xi$
plus polarization (DZP) basis set of Slater-type orbitals, taken
from the ADF database. Such a density is then used to build the
$h_{KS}$ hamiltonian (Cfr. Eq.~(\ref{eq1})). The LB94 $xc$ potential
\cite{LB94} has been used because of its correct asymptotic behavior
\cite{Mahan90}.

The fixed-nuclei photoionization calculations used an SCE expansion
up to $l_{max}^{SCE}=15$ for expanding the bound molecular and
continuum orbitals with the SCE placed on the centre of mass (C.M.)
of the molecule. $B$-spline functions of order $10$ are employed in
the calculation, and defined over a linear radial grid with a step
size of $0.2$ a.u. extending up to $R_{max}^{SCE}=20$ a.u.. The
intervals were supplemented with additional knots near the position
of the nuclei in order to gain flexibility of the basis set in the
region of the core orbitals. The maximum angular momenta
$l_{max}^{i}$ employed in the off-center expansions were
$l_{max}^{i}=1$ and $l_{max}^{i}=2$ for the hydrogens and the
heavier atoms respectively, whereas carefully selected values for
$R_{max}^{i}$ defining the off-center radial grids range from 0.75
a.u. (hydrogen sites) to 1.24 a.u. (oxygen sites).

Valence IP's of Uracil where also evaluated
with the ADF program by using the Slater's transition-state
approximation~\cite{Slater74}, a Slater-type basis set of
triple-$\xi$ plus polarization (TZP) quality and the
Becke-Perdew~\cite{Becke88,Perdew86}~\emph{xc} potential.

\section{\label{sec:section4}Results and Discussion}

\begin{figure}[p]
  \begin{center}
  \includegraphics[height=3cm]{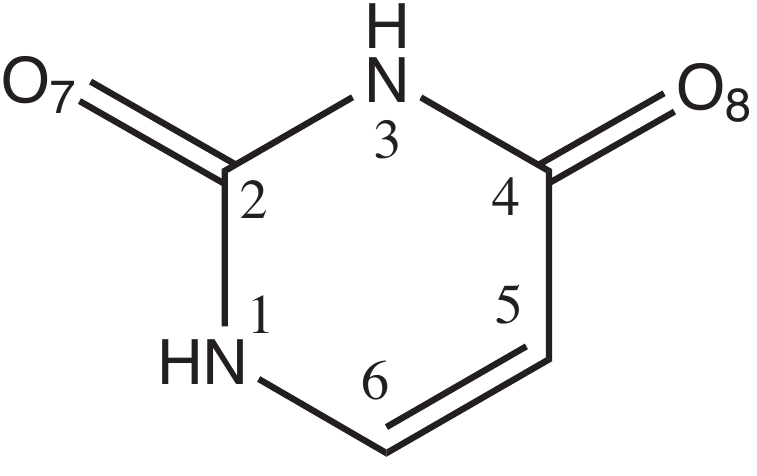}
  \end{center}
  \caption{
  Chemical structure for Uracil and numbering scheme adopted in the present work.
  }
  \label{Fig.1}
\end{figure}

In this section we will present and discuss photoionization cross
sections and asymmetry parameters profiles calculated with the LCAO
DFT scattering code for selected orbital ionizations. Specifically
the K $1s$ ionizations of Oxygen Nitrogen and Carbon atoms, and the
outer-valence ionizations giving rise to the first five resolved
bands of the photoelectron spectrum~\cite{ODonnell80, Urano89,
Kubota96}. Although results have been obtained also for
inner-valence orbital ionizations we will not attempt to discuss and
rationalize them here, since single-electron approximation based
theories are well-known to loose of their predictive power in the
inner-valence region of the spectrum. (magari mettere su un
documento EPAPS?) Figure~\ref{Fig.1} reports the chemical structure
of Uracil and the numbering scheme adopted in the present work,
while the experimental IPs~\cite{ODonnell80, Urano89, Kubota96} are
reported in Table I, together with theoretical values taken from the
literature~\cite{Dolgounitcheva00}, and results of DFT bound-state
calculations using two different \emph{xc} potentials and
ground-state (core ionizations) or transition-state (valence
ionizations) configurations. Focusing on the core IP's, fair
agreement is found between experimental~\cite{Peeling78} and
ground-state LB94 IP's, the latter being systematically
overestimating the core IP's energies by as much as $5$ eVs.
Nonetheless, the LB94 ground-state calculations are able to account
for the small differences among the various C $1s$ IP's, in other
words the chemical shifts of carbon atoms are nicely reproduced, as
usually found when applying DFT to core ionizations of large
molecules (see e.g.~\cite{Stener94}). The observed trend in the
chemical shifts is readily rationalized by considering the bond
connectivity of the carbon atoms, in particular with the more
electronegative nitrogen and oxygen centers. These results once more
underline the peculiarity of core ionization studies to provide
qualitative informations of both electronic and structural type,
pertaining to the different sites probed by core ionization.

For the five outermost valence IP's there is a broad agreement between
theoretical results and experimental data: use of the Becke-Perdew
~\cite{Becke88,Perdew86} \emph{xc} potential and the transition-state procedure
with a TZP basis set has allowed to account fairly well of relaxation
and correlation effects, giving results of comparable quality of
those obtained with the P$3$ \emph{Ab-Initio}
procedure~\cite{Dolgounitcheva00}.

\subsection{Photoelectron dynamics of core ionizations}

\begin{figure}[p]
  \begin{center}
  \includegraphics[height=12cm]{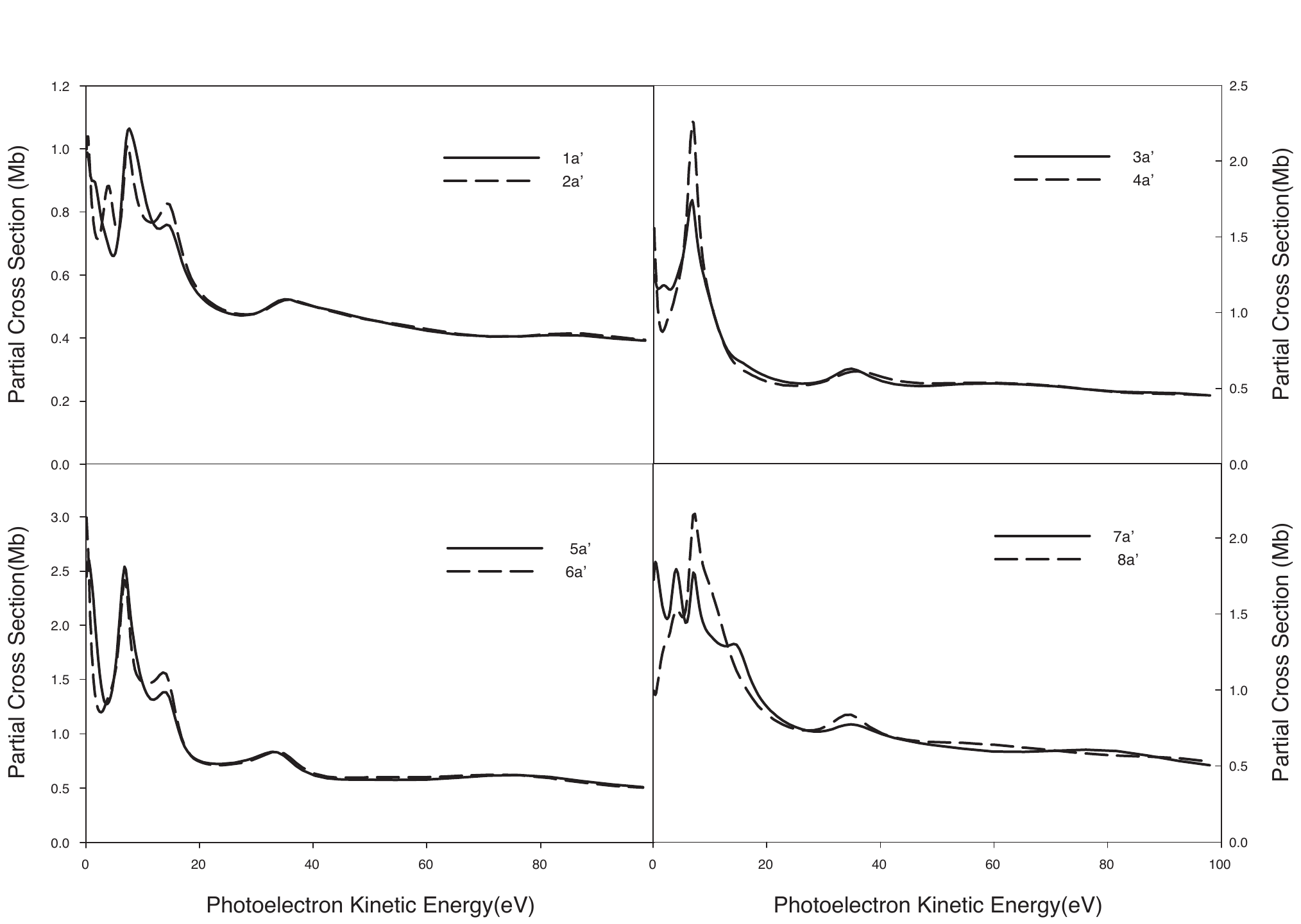}
  \end{center}
  \caption{
  LCAO DFT partial cross sections for $1s$ ionization of Oxygen (upper-left panel), Nitrogen (upper-right panel)
  and Carbon (lower panels) in the Uracil molecule.
  }
  \label{Fig.2}
\end{figure}

\begin{figure}[p]
  \begin{center}
  \includegraphics[height=12cm]{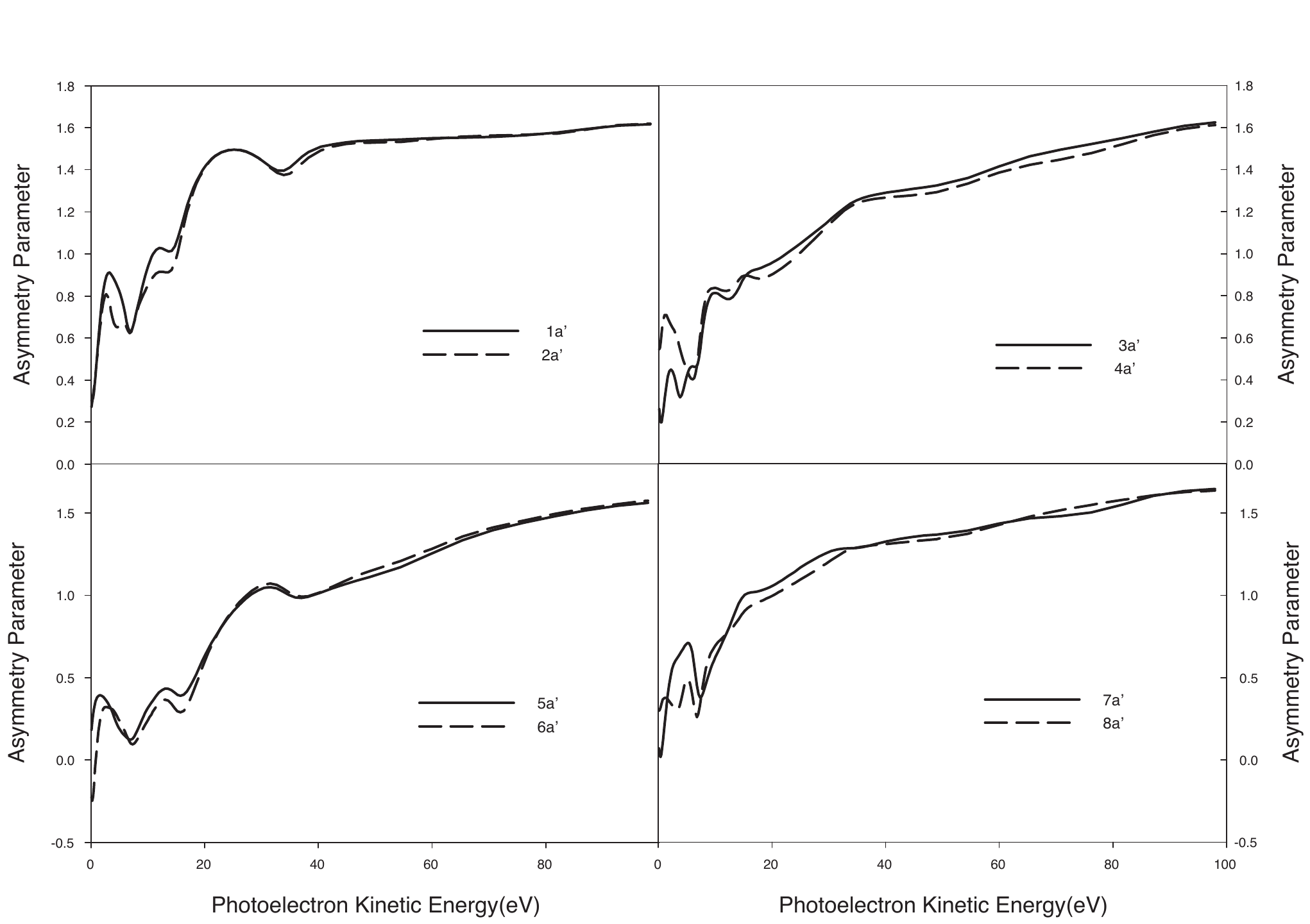}
  \end{center}
  \caption{
  LCAO DFT asymmetry parameter profiles for $1s$ ionization of Oxygen (upper-left panel), Nitrogen (upper-right panel)
  and Carbon (lower panels) in the Uracil molecule.
  }
  \label{Fig.3}
\end{figure}

The LCAO-DFT partial cross section and asymmetry parameter profiles
for the core ionizations, Oxygen $1s$ ($1a^{'}$ and $2a^{'}$
ionizations), Nitrogen $1s$ ($3a^{'}$ and $4a^{'}$ ionizations) and
Carbon $1s$ ($5a^{'}$ through $8a^{'}$ ionizations), are plotted in
Figure~\ref{Fig.2} and Figure~\ref{Fig.3} respectively. The
dynamical observables are plotted on the photoelectron kinetic
energy scale because it will prove convenient for the purpose of
identifying striking similarities in the scattering dynamics for the
various final ionic states. One readily sees that the two Oxygen
$1s$ ionization cross sections behave similarly, due to nearly
identical initial states (upper left frame of Fig.~\ref{Fig.2}). The
same is true for the Nitrogen $1s$ ionizations (upper right frame of
Figure~\ref{Fig.2}). Three low-energy shape-resonances are clearly
discernible in the O $1s$ ionization channels, a more intense one
centered at about $7.5$ eV of photoelectron kinetic energy, located
in between two less intense features and superimposed on a strong
background. A sharper prominent resonance characterize the
low-energy scattering dynamics following the N $1s$ ionizations,
with a peak energy of about $6.8$ eV. The resonant peak shows up
more clearly in the N$_{3}$ $1s^{-1}$ continua, rather than in the
N$_{1}$ one, where the cross section is more structured toward
threshold. In the lower frames of Fig. \ref{Fig.2} we have reported
the cross section profiles for the four C $1s^{-1}$ ionizations.
These are conveniently plotted in two separate groups, according to
the similarities or differences shared in their near-threshold
behavior. Thus one can easily find marked analogies in the $5a^{'}$
and $6a^{'}$ partial cross sections over the whole electron kinetic
energy range: both display a fairly strong resonant peak at about
$7$ eV and a weaker one at about $14.1$ eV, after which the two
profiles can be nearly superimposed on the scale of the Figure. The
cross section profiles for the $7a^{'}$ and $8a^{'}$ orbital
ionizations behave quite similarly, even if two separate features at
$7.1$ and $14.1$ eV in the $7a^{'}$ continua eventually coalesce
into a single broad peak in the $8a^{'}$ continua, and the two cross
section profiles behave markedly different when approaching
threshold. It is worth noting that all core ionization cross
sections show a broad hump at about 35 eV of photoelectron kinetic
energy, a feature which should be connected with inefficient
trapping of the photoelectron by high-$l$ partial waves of the
one-particle effective scattering potential. Asymmetry parameter
profiles, plotted in Fig. \ref{Fig.3}, behave similarly: marked
analogies are again found within the four groups of core
ionizations. In every case the profiles are quite structured, a
clear consequence of the wealth of resonant states that characterize
the low-energy scattering dynamics of Uracil core ionization. The
high-energy shape-resonant state, which is predicted to appear in
all the ionization continua, causes stronger modulations on the
angular distributions of the two O $1s^{-1}$ and C$_{2}$ and C$_{4}$
$1s^{-1}$ continua when compared with those leading to the remaining
core-ionized target states. Furthermore, by inspecting the partial
photoionization cross section profiles for the two dipole-allowed
continuum symmetries (not reported in the figures) it is found that
for every target ion the continuum resonant states belong to the
$a^{'}$ symmetry (\emph{vide infra}).

\subsection{Photoelectron dynamics of outer valence ionizations}

\begin{figure}[p]
  \begin{center}
  \includegraphics[height=18cm]{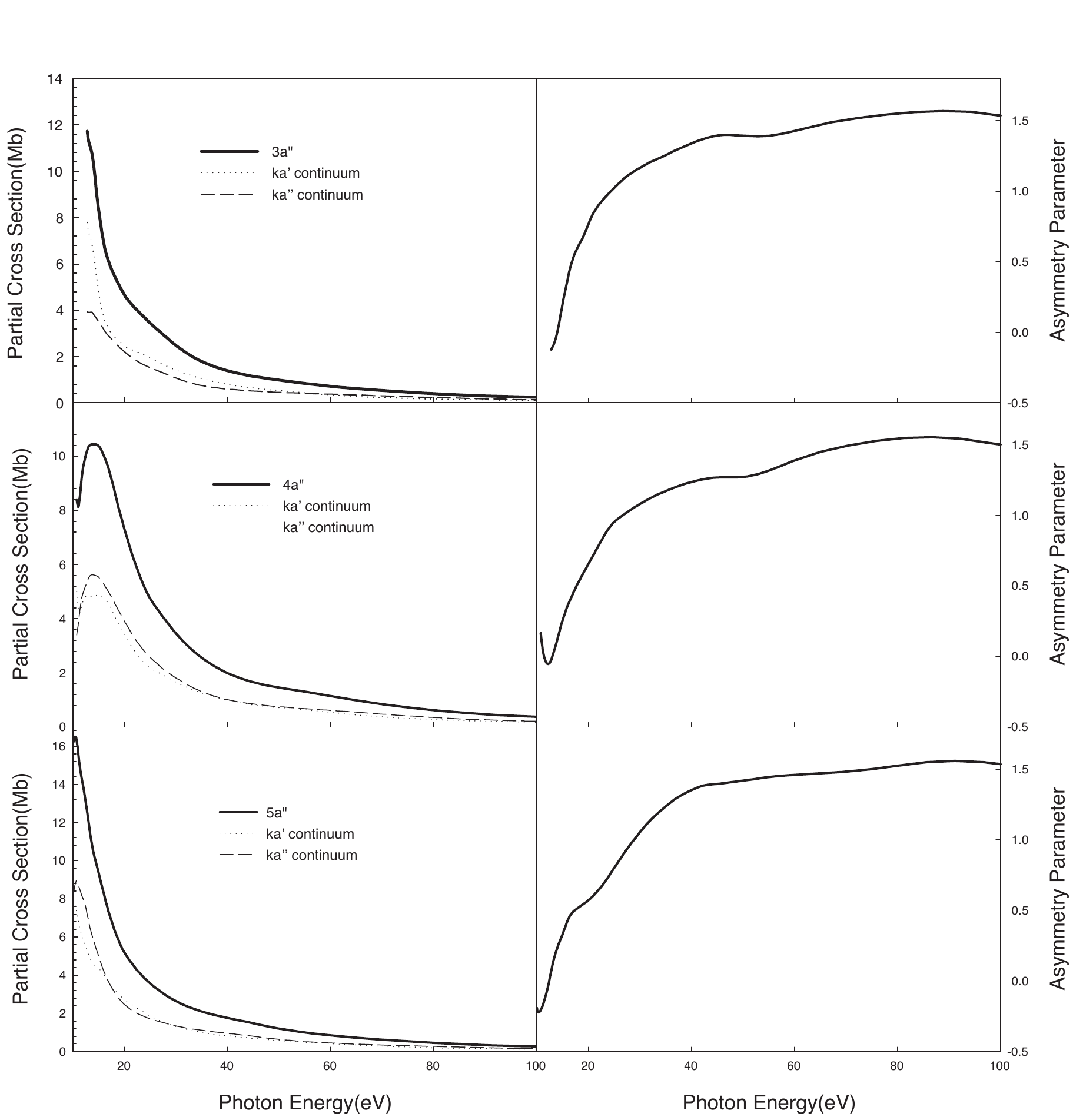}
  \end{center}
  \caption{
  LCAO DFT partial cross section and asymmetry parameter profiles for the $3a^{''}$ (upper panels), $4a^{''}$ (central panels),
  and $5a^{''}$ (lower panels) orbital ionizations of Uracil.
  }
  \label{Fig.4}
\end{figure}

\begin{figure}[p]
  \begin{center}
  \includegraphics[height=12cm]{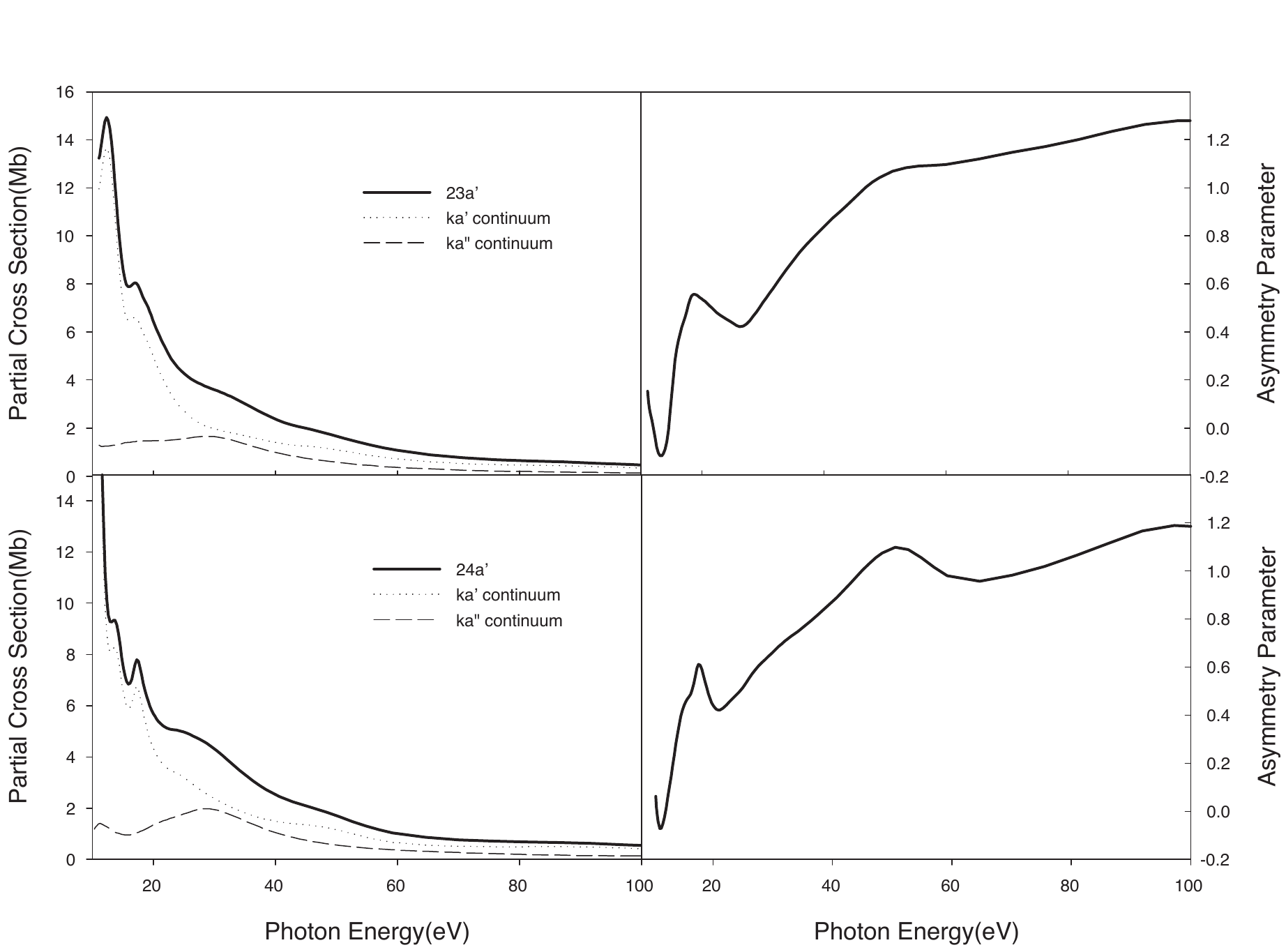}
  \end{center}
  \caption{
  LCAO DFT partial cross section and asymmetry parameter profiles for the $23a^{'}$ (upper panels), and $24a^{'}$ (lower panels)
  orbital ionizations of Uracil.
  }
  \label{Fig.5}
\end{figure}

The LCAO-DFT partial cross section and asymmetry parameter profiles
for the outermost photoelectron band orbital ionizations are plotted
in Figures~\ref{Fig.4} for the $\pi$ orbitals ($3a^{''}$, $4a^{''}$
and $5a^{''}$) and in Figure~\ref{Fig.5} for the (oxygen lone pairs)
nonbonding $n$ orbitals ($23a^{'}$ and $24a^{'}$). As anticipated,
the richness of resonant features characterizing the inner-shell
ionizations is partially lost when moving to the outer-valence
ionizations, even if strong and well defined features clearly show
up in at least two continuum channels (corresponding to the
$4a^{''}$ and $23a^{'}$ ionizations, central and upper panels of
Fig.~\ref{Fig.4} and Fig.~\ref{Fig.5} respectively). Focusing our
attention to the $\pi$ ionizations (Figure~\ref{Fig.4}) the cross
section profiles share a common behavior: a steep rise toward
threshold and a rapid monotonic decrease for higher excitation
energies. However, only in the $4a^{''}$ continua above-threshold
shape resonances are found, where both the $\epsilon a^{'}$ and
$\epsilon a^{''}$ channels are predicted to be resonant. Partial and
total cross section profiles for the photoionization out of the
$3a^{''}$ orbital rise steeply at threshold, without any noticeable
slope change, and a resonant state, albeit very close to threshold
is visible in the $\epsilon a^{''}$ continuum of the $5a^{''}$
ionization and is barely apparent in the summed profile (lower left
panel of Figure~\ref{Fig.4}). The corresponding asymmetry
parameters, also plotted in Figure~\ref{Fig.4} are almost
featureless, with the exception of that corresponding to the
$4a^{''}$ ionization which displays a near-threshold oscillation
typical of the formation of a resonant state, as discussed above.

Partial and total cross sections for the remaining two valence
ionizations considered ($23a^{'}$ and $24a^{'}$ orbitals) are
predicted to be quite more structured and are plotted in
Figure~\ref{Fig.5}. Both cross sections show strong oscillations
near threshold in the $\epsilon a^{'}$ continuum, and a rather broad
peak above $30$ eV in the $\epsilon a^{''}$ channel. Similarly, the
asymmetry parameter profiles are rather structured, both at low and
high photon energies. Table~\ref{Table.2} lists, for every orbital
ionization analyzed so far, the peak energy positions (on the
photoelectron kinetic energy scale) and the symmetries of the
resonant states identified.

\subsection{Correlation with electron-molecule scattering experiments}

As readily apparent from Table II, computed photoionization dynamics
of Uracil reveals a wealth of near-threshold structures, especially
in core ionizations. For core ionizations several resonant states
show up at roughly the same energy positions into several final
target state channels. This is the case, for example, for the
resonant $\epsilon a^{'}$ states with peak energy positions at about
$7$ eV, $14.1$ eV and about $34$ eV of photoelectron kinetic energy.
As an attempt to correlate computed photoionization resonant states
with those found in electron-molecule scattering
calculations~\cite{Grandi04, Gianturco04} we then decided to analyze
the resonant wave functions by inspecting the
~\emph{dipole-prepared} continuum wave function~\cite{Stener02}.
Therefore the continuum wavefunction corresponding to the symmetry
and peak energy positions listed in Table II have been
 transformed accordingly~\cite{Stener02} and analyzed.
Nearly all the \emph{pseudo}-bound states are diffuse in nature,
extending thorugh the whole molecular skeleton. Also no hint of
resonant states that could be correlated with the metastable anionic
state experimentally found at about 9 eV in electron-scattering
experiments were found. This is quite expected in view of the
different nature of the interaction potential in photoionization
compared to electron-neutral target collision processes. We are
therefore led to the conclusion that all relevant resonant states
characterized in low-energy electron collision
processes~\cite{Grandi04, Gianturco04}, shift below threshold in
photoionization. For the purpose of comparison with the present
study it would be interesting to investigate higher energy
electron-collision processes; modeling these high energy
interactions proves however a more difficult theoretical task since
electronic excitation channels need to be considered as well.

Finally, one last comment about our theoretical results should be
made. Since vibrational couplings are completely neglected in our
treatment of the photoionization dynamics and those invariably
shorten the life-time of metastable resonant states, a broadening of
the resonant structures will result with the consequence that the
beautiful structuring of core ionizations cross sections predicted
in the framework of the fixed-nuclei approximation will be partially
lost.

\begin{table*}[p]
\caption{\label{Table.1}Uracil Ionization Energies (eV) (for the
numbering scheme adopted, see Fig~\ref{Fig.1}).}
\begin{ruledtabular}
\begin{tabular}{ccccc}
\, &Orbital Ionization & P3 results\footnotemark[1]  & DFT results, present work\footnotemark[2] & Exp.\footnotemark[4] \\
\hline
\, & $1a^{'}$(O$_{7}1s$) & & 536.8 & 531.8 \\
\, & $2a^{'}$(O$_{8}1s$) & & 536.7 & 531.8 \\
\, & $3a^{'}$(N$_{1}1s$) & & 406.9 & 400.8 \\
\, & $4a^{'}$(N$_{3}1s$)& & 406.5 & 400.8 \\
\, & $5a^{'}$(C$_{2}1s$) & & 294.3 & 289.7 \\
\, & $6a^{'}$(C$_{4}1s$)& & 293.5 & 288.9 \\
\, & $7a^{'}$(C$_{6}1s$)& & 292.6 & 286.8 \\
\, & $8a^{'}$(C$_{5}1s$)& & 291.1 & 285.6 \\
\, & $3a^{''}$($\pi_{3}$) & 12.91  & 12.62\footnotemark[3] & 12.5-12.7 \\
\, & $23a^{'}$($n_{2}$)& 11.12  & 11.40\footnotemark[3] & 10.9-11.2 \\
\, & $4a^{''}$($\pi_{2}$)& 10.52  & 10.72\footnotemark[3] & 10.5-10.6 \\
\, & $24a^{'}$($n_{1}$) & 10.15  & 9.89\footnotemark[3] & 10.02-10.23\\
\, & $5a^{''}$($\pi_{1}$)& 9.54   & 9.63\footnotemark[3] & 9.45-9.60  \\
\end{tabular}
\end{ruledtabular}
\footnotetext[1]{Ref.~\onlinecite{Dolgounitcheva00}, obtained with a $6-311$G$^{**}$ basis set.}
\footnotetext[2]{g.s.~LB94 \emph{xc} potential results, see text for details.}
\footnotetext[3]{t.s.~BP \emph{xc} potential results, see text for details.}
\footnotetext[4]{Refs.~\onlinecite{ODonnell80, Urano89, Kubota96} and references therein.}
\end{table*}

\begin{table*}[p]
\caption{\label{Table.2}Peak energy positions (photoelectron kinetic
energy, eV) and symmetry of computed resonant states in the core and
valence photoionization from Uracil.}
\begin{ruledtabular}
\begin{tabular}{cccccc}
\, & Orbital Ionization & Peak position  & Continuum symmetry  \\
\hline
\, & $1a^{'}$ & 7.6, 14.1, 34.8  & $\epsilon a^{'}$ \\
\, & $2a^{'}$ & 4.1, 7.3, 14.1, 35.9  & $\epsilon a^{'}$  \\
\, & $3a^{'}$ & 6.8, 34.8  & $\epsilon a^{'}$  \\
\, & $4a^{'}$ & 6.8, 35.9  & $\epsilon a^{'}$  \\
\, & $5a^{'}$ & 0.5, 6.8, 14.1, 33.7 & $\epsilon a^{'}$  \\
\, & $6a^{'}$ & 6.8, 14.1, 33.7  & $\epsilon a^{'}$  \\
\, & $7a^{'}$ & 0.4, 4.1, 7.1, 14.1, 34.8 & $\epsilon a^{'}$  \\
\, & $8a^{'}$ & 4.1, 7.3, 34.8   & $\epsilon a^{'}$  \\
\, & $23a^{'}$ & 1.4, 6.0, 17.4 & $\epsilon a^{'}$, $\epsilon a^{'}$, $\epsilon a^{''}$   \\
\, & $4a^{''}$ & 4.1, 3.3  & $\epsilon a^{'}$, $\epsilon a^{''}$   \\
\, & $24a^{'}$ & 3.5, 7.1, 18.5  & $\epsilon a^{'}$, $\epsilon a^{'}$, $\epsilon a^{''}$  \\
\, & $5a^{''}$ & 1.1 & $\epsilon a^{''}$ \\
\end{tabular}
\end{ruledtabular}
\end{table*}

\section{\label{sec:section5}Conclusion}

The paper present a theoretical investigation of the photoionization
process from the biologically important RNA base Uracil. To the
author's knowledge this is the first theoretical investigation of
photoionization dynamics from the Uracil molecule to date. The
theoretical method employed adopt
 a density functional theory description of photoionization
dynamics and take advantage of a multicentric basis set of
$B$-spline functions for obtaining results that are converged within
the one-electron effective DFT hamiltonian. Several peculiar results
were found that demand for an experimental investigation and
verification, especially in core ionizations. We predict inner-shell
photoionization dynamics to be characterized by the presence of
several resonant states and theoretical results can be rationalized
as a consequence of the inherent site specificity of core ionization
processes. Computed resonant states have been classified based on
the symmetry of the resonant state and the peak energy position of
the spectral features. Correlation with computed resonant structures
found in low-energy electron-molecule collision experiments has been
attempted: we suggest that transient negative ion states
characterized in low-energy scattering experiments are not visible
in photoionization but shifted in the discrete region of the
spectrum. It is believed that the present theoretical results could
be valuable in guiding and assisting future experimental work on
this biologically important molecule.

\section{Acknowledgments}
One of us (D. T.) would like to acknowledge CNR-INFM Democritos for
a post-doctoral research fellowship.


\end{document}